\documentclass[journal]{IEEEtran}

\usepackage{cite}
\usepackage{graphicx}
\usepackage{amsmath,amssymb}
\usepackage{booktabs}
\usepackage{multirow}
\usepackage{siunitx}
\usepackage{url}
\usepackage[hidelinks,hypertexnames=false]{hyperref}
\usepackage{balance}

\usepackage{algorithm}
\usepackage{algorithmic}

\usepackage{tikz}
\usetikzlibrary{arrows.meta,positioning,shapes.geometric}

\newcommand{\trpms}{\textit{IEEE Transactions on Radiation and Plasma Medical Sciences}}
\newcommand{\dt}{digital twin}

\newcommand{\openkbp}{OpenKBP}
\newcommand{\openkbpfull}{Open-access Knowledge-Based Planning (\openkbp{})}
\newcommand{\kbp}{knowledge-based planning}

\begin{document}

\title{Toward Actionable Digital Twins for Radiation-Based Imaging and Therapy:
Mathematical Formulation, Modular Workflow, and an OpenKBP-Based Dose-Surrogate Prototype}

\author{Hsin-Hsiung~Huang,~\IEEEmembership{Member,~IEEE,}
        and~Bulent~Soykan,~\IEEEmembership{Member,~IEEE}%
\thanks{This work did not involve human subjects or animals in its research. H.-H. Huang was supported in part by the U.S. National Science Foundation under Grants DMS-1924792 and DMS-2318925.}%
\thanks{Hsin-Hsiung Huang is with the School of Data, Mathematical, and Statistical Sciences, University of Central Florida, Orlando, FL 32816 USA (e-mail: hsin.huang@ucf.edu).}%
\thanks{Bulent Soykan is with the Institute for Simulation and Training, University of Central Florida, Orlando, FL 32826 USA (e-mail: Bulent.Soykan@ucf.edu).}%
}

\markboth{\trpms, Vol.~XX, No.~X, Month~2026}%
{Huang and Soykan: Toward Actionable Digital Twins for Radiation-Based Imaging and Therapy}

\maketitle

\begin{abstract}
Digital twins for radiation-based imaging and therapy are most useful when they assimilate patient data, quantify predictive uncertainty, and support clinically constrained decisions. This paper presents a modular framework for actionable digital twins in radiation-based imaging and therapy and instantiates its reproducible open-data component using the \openkbpfull{} benchmark. The framework couples PatientData, Model, Solver, Calibration, and Decision modules and formalizes latent-state updating, uncertainty propagation, and chance-constrained action selection. As an initial implementation, we build a GPU-ready PyTorch/MONAI reimplementation of the \openkbp{} starter pipeline: an 11-channel, 19.2M-parameter 3D U-Net trained with a masked loss over the feasible region and equipped with Monte Carlo dropout for voxel-wise epistemic uncertainty. To emulate the update loop on a static benchmark, we introduce decoder-only proxy recalibration and illustrate uncertainty-aware virtual-therapy evaluation using DVH-based and biological utilities. A complete three-fraction loop including recalibration, Monte Carlo inference, and spatial optimization executes in 10.3~s. On the 100-patient test set, the model achieved mean dose and DVH scores of 2.65 and 1.82~Gy, respectively, with 0.58~s mean inference time per patient. The \openkbp{} case study thus serves as a reproducible test bed for dose prediction, uncertainty propagation, and proxy closed-loop adaptation, while future institutional studies will address longitudinal calibration with delivered-dose logs and repeat imaging.
\end{abstract}


\section{Introduction}
The idea of a \dt\ emerged in aerospace and engineered systems as a virtual counterpart that remains linked to its physical system through data, modeling, and prediction~\cite{Glaessgen2012DigitalTwin}. In medicine, the concept is especially compelling because patient care already generates a rich digital record of images, treatment plans, delivery logs, and clinical outcomes. Radiation-based imaging and therapy are particularly well suited to this paradigm because they combine repeated measurement, model-based prediction, and decisions that must be revised under strict safety constraints. In this setting, a useful \dt\ is not merely a static simulator. It must assimilate new patient information over time, represent uncertainty in its predictions, and support actions such as plan evaluation, adaptation, or treatment selection.

Radiation oncology already provides two ingredients needed for such systems. First, deep learning has advanced dose prediction and \kbp, with the \openkbpfull{} AAPM Grand Challenge providing a standardized benchmark for volumetric dose surrogates and DVH-based evaluation~\cite{Babier2021OpenKBP,OpenKBPRepo}. Second, digital-twin thinking has matured in oncology and medical imaging, including recent patient-specific forecasting frameworks and MRI-based response models~\cite{Wu2025MRIDigitalTwin,Kapteyn2025TumorTwin}. Together, these developments suggest a timely opportunity to connect reproducible AI-based dose modeling with the updating and decision logic expected of an actionable \dt.

That connection remains incomplete. Most published dose-prediction models are still one-shot feed-forward predictors. They estimate a reference dose distribution from anatomy and structure masks, but they do not explicitly maintain a patient state, incorporate new observations during treatment, or express uncertainty in a way that can guide decisions. A clinically meaningful radiation-therapy \dt\ therefore requires more than predictive accuracy alone. It requires a formal updating mechanism, uncertainty propagation from voxels to clinically interpretable summaries, and a decision layer that can compare candidate actions under safety and deliverability constraints. Open datasets are indispensable for reproducibility, yet they rarely contain the longitudinal measurements needed for a full bidirectional clinical twin. This paper addresses that tension by using \openkbp{} for reproducible dose-surrogate development, uncertainty-aware evaluation, and proxy sequential updating, while reserving true longitudinal validation for future institutional studies with delivered-dose logs and repeat imaging.

Within that scope, we develop an actionable digital-twin framework organized around PatientData, Model, Solver, Calibration, and Decision. The open-data implementation uses an \openkbp{}-style 3D U-Net trained with a masked loss over the feasible region, stochastic inference through Monte Carlo dropout for voxel-wise epistemic uncertainty, and a proxy sequential calibration step based on decoder-only transfer learning. These components are coupled to a virtual-therapy decision module that evaluates candidate actions using DVH-based summaries and uncertainty-aware biological utilities. The result is both a mathematical formulation of an actionable \dt\ and a reproducible prototype showing how uncertainty and updating can be embedded in a benchmark dose-prediction workflow.

\section{Background and Related Work}
\subsection{Updateable patient models and modular twins}
Across engineering and biomedical settings, digital twins are virtual counterparts linked to their physical systems through data assimilation, prediction, and decision support~\cite{Glaessgen2012DigitalTwin}. In oncology, recent examples span MRI-based patient-specific digital twins for chemotherapy optimization~\cite{Wu2025MRIDigitalTwin} and modular frameworks such as TumorTwin~\cite{Kapteyn2025TumorTwin}. These lines of work collectively motivate the present focus on updating, uncertainty-aware prediction, and actionability.

\subsection{Radiation-therapy surrogates and open benchmarks}
\openkbp{} enables fair comparison of dose-prediction methods for \kbp{} using standardized voxel-wise and DVH-based metrics~\cite{Babier2021OpenKBP,OpenKBPRepo}. Attention-aware 3D U-Net variants have also been evaluated on the benchmark and show that architectural refinements can improve emphasis on clinically relevant anatomy in head-and-neck IMRT dose prediction~\cite{Osman2022AttentionUNet}. Our work builds on this benchmark tradition but moves beyond static prediction by integrating uncertainty, proxy recalibration, and a decision layer.

\section{Model of an Actionable Digital Twin}
\subsection{Latent state, observations, and updating}
Let $x_t$ denote the latent patient state at time $t$, including anatomy, targets and organs-at-risk, accumulated dose summaries, and latent response or toxicity parameters. Let $y_t$ denote observations, including CT or CBCT, delivered dose logs, biomarkers, and outcomes, and let $u_t$ denote clinical actions such as plan parameters, adaptation triggers, and imaging schedules. A generic state-space model is
\begin{align}
x_{t+1} &= f(x_t,u_t;\theta) + w_t, \label{eq:state}\\
y_t &= h(x_t;\phi) + v_t, \label{eq:obs}
\end{align}
where $\theta$ denotes transition or calibration parameters, $\phi$ denotes observation-model parameters, and $w_t$ and $v_t$ represent process and observation uncertainty. For notational simplicity, we treat $\phi$ as fixed or pre-specified during online updating and maintain the belief state over $(x_t,\theta)$. Accordingly,
\begin{align}
b_t(x,\theta) \triangleq p(x_t=x,\theta_t=\theta \mid y_{1:t},u_{1:t-1};\phi). \label{eq:belief}
\end{align}
The Bayesian filtering update is
\begin{align}
b_t(x_t,\theta) &\propto p(y_t\mid x_t;\phi) \nonumber\\
&\quad \times \int p(x_t\mid x_{t-1},u_{t-1};\theta)\, b_{t-1}(x_{t-1},\theta)\, dx_{t-1}. \label{eq:bayes}
\end{align}
In practice, \eqref{eq:bayes} can be approximated by ensemble, particle, or maximum a posteriori (MAP) updates. A generic MAP calibration step is
\begin{align}
(\hat x_t,\hat\theta_t) &= \arg\min_{x,\theta}\; \|y_t-h(x;\phi)\|_{\Sigma_v^{-1}}^2 \nonumber\\
&\quad + \|x-f(\hat x_{t-1},u_{t-1};\theta)\|_{\Sigma_w^{-1}}^2 + R(\theta), \label{eq:map}
\end{align}
where $\Sigma_v$ and $\Sigma_w$ are error covariances and $R(\theta)$ encodes priors or stability.

\subsection{Dose surrogate, uncertainty, and DVH metrics}
For radiation therapy, the predictive target is the 3D dose distribution $d(\mathbf{r})$ over voxels $\mathbf{r}$. Let $I_t$ denote the planning CT and let $S_t$ denote ROI masks. A dose surrogate is defined as
\begin{align}
\hat d_t = g_{\psi}(I_t,S_t), \label{eq:surrogate}
\end{align}
where $\psi$ denotes neural-network parameters. For \openkbp, $g_{\psi}$ is trained to predict the reference plan dose using the masked objective
\begin{align}
\psi^\star = \arg\min_{\psi}\sum_{i=1}^{N} \left\| M_i \odot \big(g_{\psi}(I_i,S_i)-d_i\big) \right\|_1, \label{eq:train}
\end{align}
where $M_i$ is the feasible dose mask~\cite{Babier2021OpenKBP,OpenKBPRepo}. Predictive uncertainty is approximated by stochastic inference,
\begin{align}
\hat d_t^{(k)} = g_{\psi}(I_t,S_t;\xi_k), \qquad k=1,\ldots,K, \label{eq:mc}
\end{align}
with predictive mean and voxel-wise variance
\begin{align}
\bar d_t &= \frac{1}{K}\sum_{k=1}^{K}\hat d_t^{(k)}, \\
\sigma_t^2(\mathbf{r}) &= \frac{1}{K-1}\sum_{k=1}^{K}\left(\hat d_t^{(k)}(\mathbf{r})-\bar d_t(\mathbf{r})\right)^2, \label{eq:uq}
\end{align}
where Monte Carlo dropout serves as an efficient approximate Bayesian mechanism for epistemic uncertainty extraction~\cite{Gal2016Dropout}. Let $\Omega_r$ denote the voxel set for ROI $r$ and let $\mathbf{m}(d)$ collect DVH-derived metrics. The benchmark-aligned dose and DVH scores are
\begin{align}
S_{\text{dose}}(\hat d,d) &= \frac{1}{|\Omega|}\sum_{\mathbf{r}\in\Omega}\left|\hat d(\mathbf{r})-d(\mathbf{r})\right|, \label{eq:dosescore}\\
S_{\text{DVH}}(\hat d,d) &= \frac{1}{L}\sum_{\ell=1}^{L}\left|\mathcal{M}_{\ell}(\hat d)-\mathcal{M}_{\ell}(d)\right|, \label{eq:dvhscore}
\end{align}
where $\mathcal{M}_{\ell}$ enumerates the benchmark DVH summaries~\cite{Babier2021OpenKBP}.

\subsection{Uncertainty-aware decision-making}
At decision time $t$, the \dt\ recommends an action by solving
\begin{align}
u_t^\star = \arg\min_{u\in\mathcal{U}_t}\; &\mathbb{E}_{b_t}\!\left[C(x_t,u)\right] \nonumber\\
\text{s.t.}\; &\mathbb{P}_{b_t}\!\left(g_j(x_t,u)\le 0\right) \ge 1-\alpha_j,\ \forall j, \label{eq:chance}
\end{align}
where $C(\cdot)$ captures coverage-sparing tradeoffs, $g_j(\cdot)$ are safety and deliverability constraints, and $\alpha_j$ are acceptable violation probabilities. Using samples from \eqref{eq:mc}, the decision problem can be approximated as
\begin{align}
u_t^\star \approx \arg\min_{u\in\mathcal{U}_t}\; &\frac{1}{K}\sum_{k=1}^{K} C\!\left(\mathbf{m}(\hat d_t^{(k,u)})\right) \nonumber\\
\text{s.t.}\; &\frac{1}{K}\sum_{k=1}^{K}\mathbf{1}\!\left[g_j(\hat d_t^{(k,u)})\le 0\right] \ge 1-\alpha_j. \label{eq:samplechance}
\end{align}
For \openkbp, $u$ may be instantiated as a plan-library selector, a spatial modulation mask, or a proxy adaptation used for virtual-therapy evaluation.

\section{Modular Architecture and Algorithm}
\subsection{Modules, architecture, and workflow}
\begin{figure}[t]
\centering
\resizebox{\columnwidth}{!}{%
\begin{tikzpicture}[
  font=\scriptsize,
  >=Latex,
  main/.style={draw, rounded corners, align=center, text width=3.1cm, minimum height=0.82cm},
  side/.style={draw, rounded corners, align=center, text width=2.35cm, minimum height=0.72cm},
  line/.style={->, thick},
  dline/.style={->, thick, dashed}
]
\node[main] (phys) {Clinical system\\patient + delivery};
\node[main, below=7mm of phys] (ingest) {Ingest + harmonize\\imaging, RT objects, logs};
\node[main, below=7mm of ingest] (pdata) {PatientData\\digital thread + provenance};
\node[main, below=7mm of pdata] (model) {Twin model\\physics + learned};
\node[main, below=7mm of model] (solver) {Solver\\forward prediction + sensitivities};
\node[main, below=7mm of solver] (pred) {Predictive distribution\\dose, DVH, risk + UQ};
\node[main, below=7mm of pred] (decision) {Decision\\utility + constraints};
\node[main, below=7mm of decision] (action) {Action $u_t$\\plan, adapt, schedule};
\node[side, right=8mm of pdata] (qa) {QA + clinician review\\validation checks};
\node[side, right=8mm of model] (calib) {Calibration\\update / inverse fit};
\draw[line] (phys) -- (ingest);
\draw[line] (ingest) -- (pdata);
\draw[line] (pdata) -- (model);
\draw[line] (model) -- (solver);
\draw[line] (solver) -- (pred);
\draw[line] (pred) -- (decision);
\draw[line] (decision) -- (action);
\draw[line, bend left=35] (action.west) to (phys.west);
\draw[dline] (pdata.east) -- (qa.west);
\draw[dline] (pred.east) -- (qa.south west);
\draw[dline] (pdata.south east) -- (calib.north west);
\draw[dline] (solver.east) -- (calib.south west);
\draw[dline] (calib.west) -- (model.east);
\end{tikzpicture}}
\caption{Actionable \dt\ architecture for radiation-based imaging and therapy. Measurements update PatientData, calibration updates model state or parameters, the solver produces predictive distributions with uncertainty, and the decision module selects constrained actions for the clinical system.}
\label{fig:arch}
\end{figure}
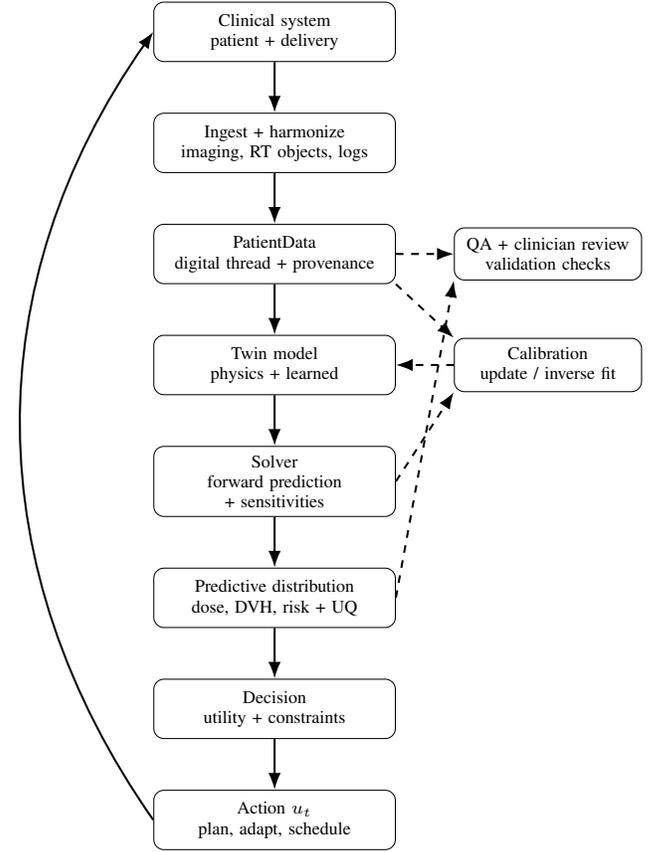

Table~\ref{tab:mapping} maps a modular \dt\ workflow to radiation-based imaging and therapy, and Fig.~\ref{fig:arch} shows the corresponding closed-loop architecture. The separation between PatientData, Model, Solver, Calibration, and Decision mirrors TumorTwin's modular philosophy while focusing on RT objects, DVH-aware objectives, and benchmark-aligned evaluation~\cite{Kapteyn2025TumorTwin}.

\begin{table}[t]
\footnotesize
\centering
\caption{Module mapping from a modular \dt\ workflow to radiation-based imaging and therapy.}
\label{tab:mapping}
\begin{tabular}{@{}p{1.8cm}p{5.5cm}@{}}
\toprule
\textbf{Module} & \textbf{Radiation-based imaging and therapy instantiation} \\
\midrule
PatientData & Harmonized imaging, structures, plan goals, dose distributions, delivered logs, metadata, and outcomes. \\
Model & Hybrid physics and learned models, including dose surrogate, accumulated dose, response, and toxicity models. \\
Solver & Forward prediction for dose and downstream summaries, with sensitivities when needed for calibration and decision. \\
Calibration & Sequential updating using new measurements, implemented as inverse problems or filtering as in \eqref{eq:bayes}-\eqref{eq:map}. \\
Decision & Constrained protocol selection using predictive distributions, utilities, and safety constraints as in \eqref{eq:chance}-\eqref{eq:samplechance}. \\
\bottomrule
\end{tabular}
\end{table}

Algorithm~\ref{alg:workflow} summarizes the hybrid-data workflow, separating reproducible open-benchmark training from patient-specific updating and decision-making. The structure is motivated by modular digital-twin workflows such as TumorTwin but is instantiated here for radiation-therapy dose prediction and uncertainty-aware adaptation~\cite{Kapteyn2025TumorTwin}.

\begin{algorithm}[t]
\footnotesize
\caption{Hybrid-data actionable \dt\ workflow}
\label{alg:workflow}
\begin{algorithmic}[1]
\STATE \textbf{Input:} open benchmark data $\mathcal{D}_{\text{open}}$, patient measurements $\{y_t\}$, candidate actions $\mathcal{U}_t$, constraints $g_j$, violation levels $\alpha_j$, stochastic sample size $K$.
\STATE Build PatientData objects for $\mathcal{D}_{\text{open}}$ and train the dose surrogate by minimizing \eqref{eq:train}; validate with \eqref{eq:dosescore}-\eqref{eq:dvhscore}.
\STATE Initialize patient-specific PatientData and belief $b_0(x,\theta)$.
\FOR{$t=1,2,\ldots,T$}
  \STATE Update PatientData with new measurement $y_t$.
  \STATE Calibrate via approximate filtering \eqref{eq:bayes} or MAP updating \eqref{eq:map}.
  \STATE Generate $K$ stochastic dose samples using \eqref{eq:mc} and compute uncertainty via \eqref{eq:uq}.
  \STATE Compute DVH summaries and evaluate candidate actions with sample-based chance constraints \eqref{eq:samplechance}.
  \STATE Select $u_t^\star$, deliver it, and log the resulting summaries for the next update.
\ENDFOR
\STATE \textbf{Output:} updated beliefs, predictive distributions, selected actions, and constraint-satisfaction statistics.
\end{algorithmic}
\end{algorithm}

\subsection{Positioning relative to prior work}
Table~\ref{tab:tumortwin} compares the present radiation-therapy prototype with TumorTwin at the framework level. Relative to TumorTwin, the novelty here is the coupling of a modular digital-twin philosophy with an explicit dose-surrogate component trained and evaluated on \openkbp{} using standardized dose and DVH metrics. Relative to stand-alone dose-prediction pipelines, the novelty lies in embedding the surrogate inside an updateable loop with calibration, uncertainty propagation, and an explicit decision module. The evaluation strategy is also distinct: open benchmarking anchors reproducible model development, whereas future institutional cohorts will test delivered-dose assimilation and translational actionability under longitudinal observation streams.

\begin{table*}[t]
\footnotesize
\centering
\caption{Framework-level comparison between TumorTwin and the proposed radiation-therapy digital-twin (RT-DT) prototype.}
\label{tab:tumortwin}
\begin{tabular}{@{}p{2.35cm}p{5.05cm}p{6.15cm}@{}}
\toprule
\textbf{Aspect} & \textbf{TumorTwin} & \textbf{Proposed RT-DT prototype} \\
\midrule
Primary domain & General oncology digital twins with mechanistic tumor-growth and response models & External-beam radiation therapy with CT/ROI-driven volumetric dose-surrogate modeling \\
Primary observations & Longitudinal imaging and tumor-response measurements & Planning CT, structure masks, feasible dose mask, reference plan dose, and proxy fraction-level dose summaries \\
Calibration target & Patient-specific mechanistic parameters and latent tumor state & Dose-surrogate state or parameters via proxy sequential calibration, with encoder freezing and decoder adaptation \\
Decision output & Alternative-treatment forecasting and optimization & Uncertainty-aware plan selection or spatial modulation under RT-specific constraints \\
Time horizon & Medium- to long-term response forecasting & Short-term fractionated adaptation and dose-evaluation loop \\
Computational profile & Numerical solvers, gradients, and model-based calibration & Rapid deep-learning inference with Monte Carlo dropout and lightweight transfer-learning updates \\
Primary evaluation & Demonstration workflows and response forecasting & Dose score, DVH score, uncertainty-aware DVHs, and proxy adaptation trajectories \\
\bottomrule
\end{tabular}
\end{table*}

\section{Open Benchmark Case Study and Results}
\subsection{Configuration and study design}
For \openkbp{}, PatientData contains CT, ROI masks, the feasible dose mask, voxel dimensions, and the reference dose distribution. The implemented surrogate is a PyTorch/MONAI reimplementation of the \openkbp{} starter framework, configured as an 11-channel 3D U-Net with approximately 19.2M trainable parameters. The design preserves the starter benchmark's masked-loss formulation while upgrading the pipeline for uncertainty-aware inference. Specifically, dropout is retained in the bottleneck and decoder at test time so that repeated stochastic forward passes produce the ensemble in \eqref{eq:mc}-\eqref{eq:uq}~\cite{Gal2016Dropout}. Attention-aware 3D U-Net models remain relevant literature comparators and future extensions, but they are not the primary implemented architecture reported here~\cite{Osman2022AttentionUNet}.

Because \openkbp{} does not include delivered-dose logs or repeat imaging, sequential updating is emulated through a lightweight proxy update: the encoder is frozen to preserve anatomy-dependent features, while the decoder is adapted to dose-related summary measurements that mimic early-fraction QA signals. The decision module ranks candidate proxy actions such as spatial masks or constrained scaling perturbations using the schematic biological utility $\mathrm{TCP}-\lambda\,\mathrm{NTCP}-\gamma\,U_t$, where $U_t$ is an uncertainty penalty derived from $\sigma(\mathbf{x})$ or an ROI-aggregated summary. The primary comparisons are deterministic versus stochastic prediction, static versus proxy-recalibrated prediction, and risk-neutral versus uncertainty-penalized decision policies. Benchmark-aligned evaluation is based on the standardized dose and DVH scores, supplemented by runtime, uncertainty-aware DVHs, and illustrative decision-trajectory summaries. Consistent with TG-166, TCP/NTCP are used here as comparative utilities rather than calibrated clinical outcome probabilities~\cite{Li2012TG166}.

\subsection{Implementation stability and benchmark performance}
The current implementation verifies that the proposed infrastructure is operational and numerically stable. In a deliberate overfitting and bounds-validation test, the masked L1 training objective decreased substantially from an initial value of approximately 28.5~Gy, confirming correct data loading, feasible-mask handling, and gradient propagation through the 3D dose-surrogate pipeline. At untrained initialization, the network produced nonnegative outputs in the range $[0.000,\,3.034]$~Gy, indicating well-behaved initialization before full training. In addition, a complete three-fraction proxy virtual-therapy loop, including sequential decoder calibration, Monte Carlo inference, and spatial optimization, executed in 10.3~s on the current hardware configuration. Together, these results establish a functioning end-to-end platform for controlled uncertainty-aware inference and proxy adaptation experiments.

Following full convergence on the GPU cluster, cohort-wide performance was evaluated on the 100-patient \openkbp{} test set. As summarized in Table~\ref{tab:benchmark_results}, the model achieved mean dose and DVH scores of $2.65 \pm 0.42$~Gy and $1.82 \pm 0.55$~Gy, respectively, with mean inference time $0.58 \pm 0.05$~s per patient. These values indicate that the present surrogate attains a practical baseline level of benchmark accuracy while enabling the stochastic inference and proxy recalibration capabilities required by the broader digital-twin formulation.

\begin{table}[t]
\footnotesize
\centering
\caption{Cohort-wide \openkbp{} test-set benchmark metrics for the present model (100 patients).}
\label{tab:benchmark_results}
\begin{tabular}{@{}lcc@{}}
\toprule
\textbf{Metric} & \textbf{Mean} & \textbf{Std. dev.} \\
\midrule
Dose score [Gy] & 2.65 & 0.42 \\
DVH score [Gy] & 1.82 & 0.55 \\
Inference time / patient [s] & 0.58 & 0.05 \\
\bottomrule
\end{tabular}
\end{table}

\subsection{Spatial and DVH uncertainty diagnostics}
\begin{figure*}[t]
\centering
\includegraphics[width=\textwidth]{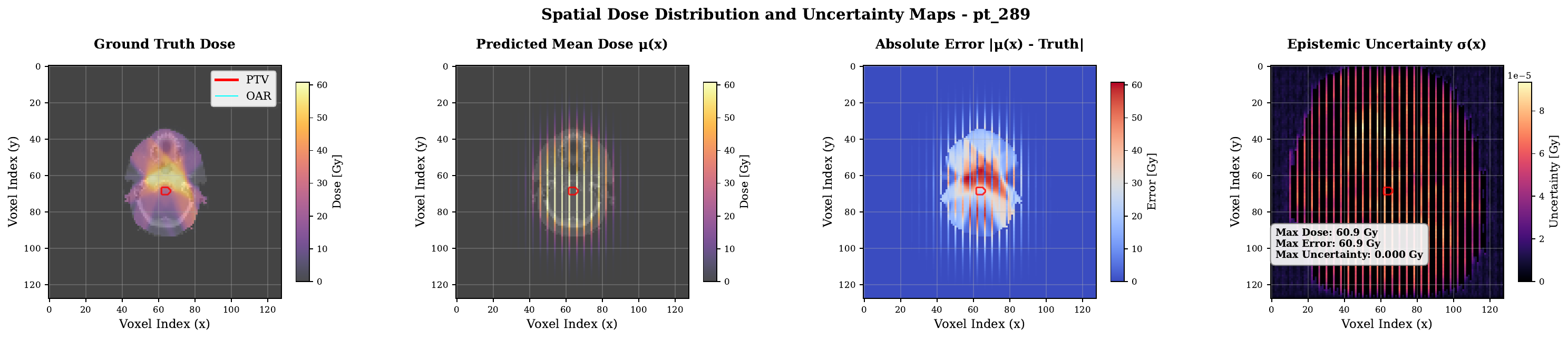}
\caption{Representative axial slice for \openkbp{} case pt\_289: reference dose, predicted mean dose $\mu(\mathbf{x})$, absolute error $|\mu(\mathbf{x})-d(\mathbf{x})|$, and voxel-wise epistemic uncertainty $\sigma(\mathbf{x})$ estimated by Monte Carlo dropout.}
\label{fig:spatial_maps}
\end{figure*}

Figure~\ref{fig:spatial_maps} shows a representative axial slice from case pt\_289 with the reference dose, predictive mean dose, absolute error, and voxel-wise epistemic uncertainty estimated by Monte Carlo dropout. The visualization localizes deviations near steep dose gradients and target/OAR interfaces, which are regions of practical importance in radiotherapy dose-prediction studies~\cite{Babier2021OpenKBP,Gal2016Dropout}. In the current exported slice, the uncertainty values are very small, so the uncertainty panel should be interpreted as a qualitative diagnostic of the reporting pipeline rather than as a standalone calibration result. The underlying uncertainty quantity is nonnegative by definition, even though the plotting routine uses a symmetric color scale.

\begin{figure}[t]
\centering
\includegraphics[width=\columnwidth]{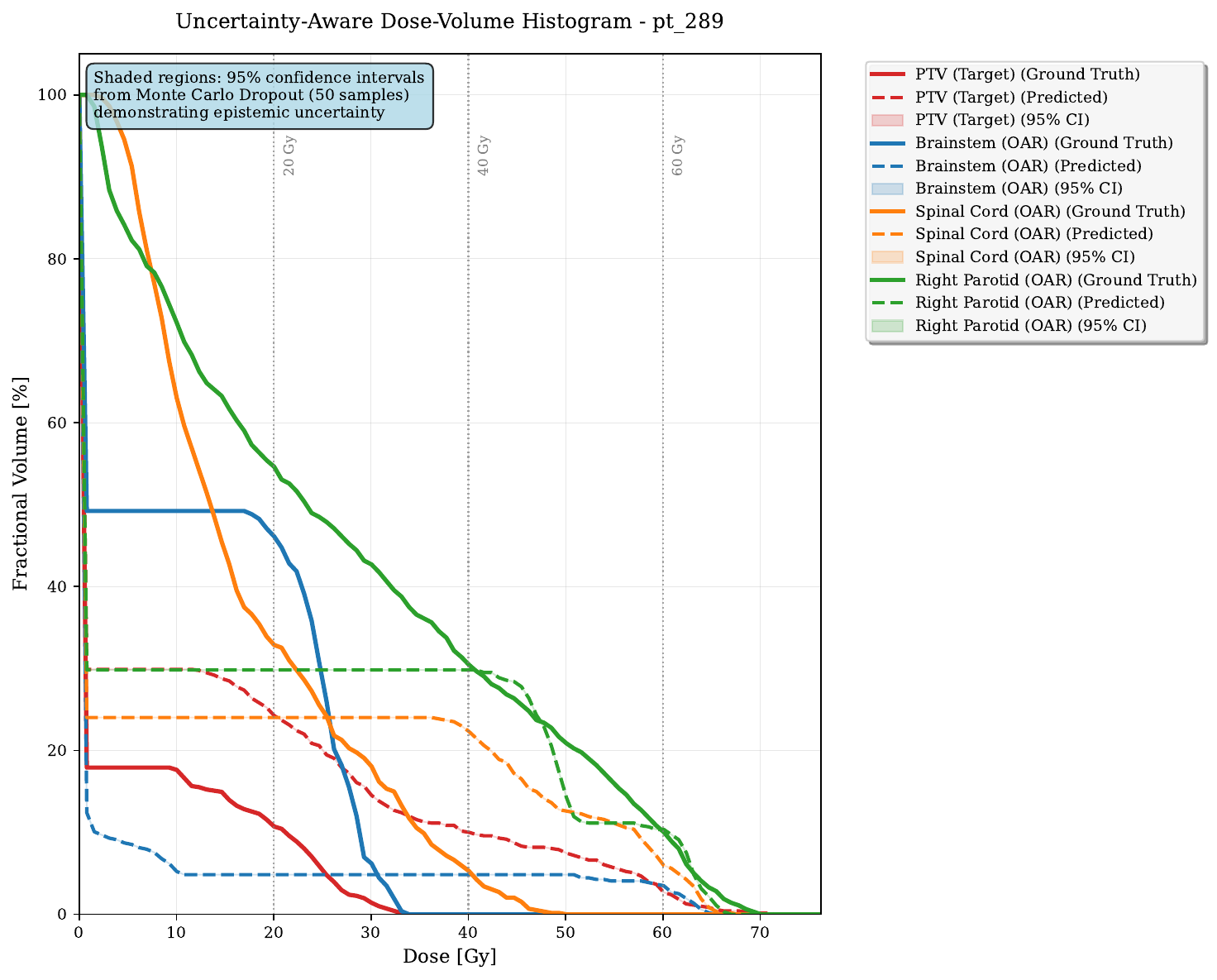}
\caption{Uncertainty-aware DVHs for case pt\_289. Solid curves denote reference DVHs, dashed curves denote predictive-mean DVHs, and shaded regions indicate pointwise 95\% predictive intervals from the Monte Carlo dropout ensemble.}
\label{fig:dvh_uq}
\end{figure}

To connect voxel-wise uncertainty to clinically interpreted plan summaries, the predictive ensemble is propagated to ROI-level DVHs, yielding pointwise 95\% predictive intervals for the PTV and representative OARs, as shown in Fig.~\ref{fig:dvh_uq}. This presentation is consistent with radiotherapy practice, where DVH-based summaries remain central to plan evaluation and uncertainty in dose should be reflected in downstream plan-quality metrics rather than reported only at the voxel level~\cite{Wahl2020DVH,Li2012TG166}. For the illustrated case, the DVH bands are narrow, which is consistent with the low voxel-wise Monte Carlo dropout spread observed in Fig.~\ref{fig:spatial_maps}. At the same time, the gap between reference and predicted DVH curves highlights that uncertainty width and predictive accuracy are distinct properties; narrow predictive intervals do not by themselves guarantee accurate dose reconstruction.

\subsection{Illustrative proxy adaptation trajectory}
\begin{figure*}[t]
\centering
\includegraphics[width=0.97\textwidth,trim=0 8 0 70,clip]{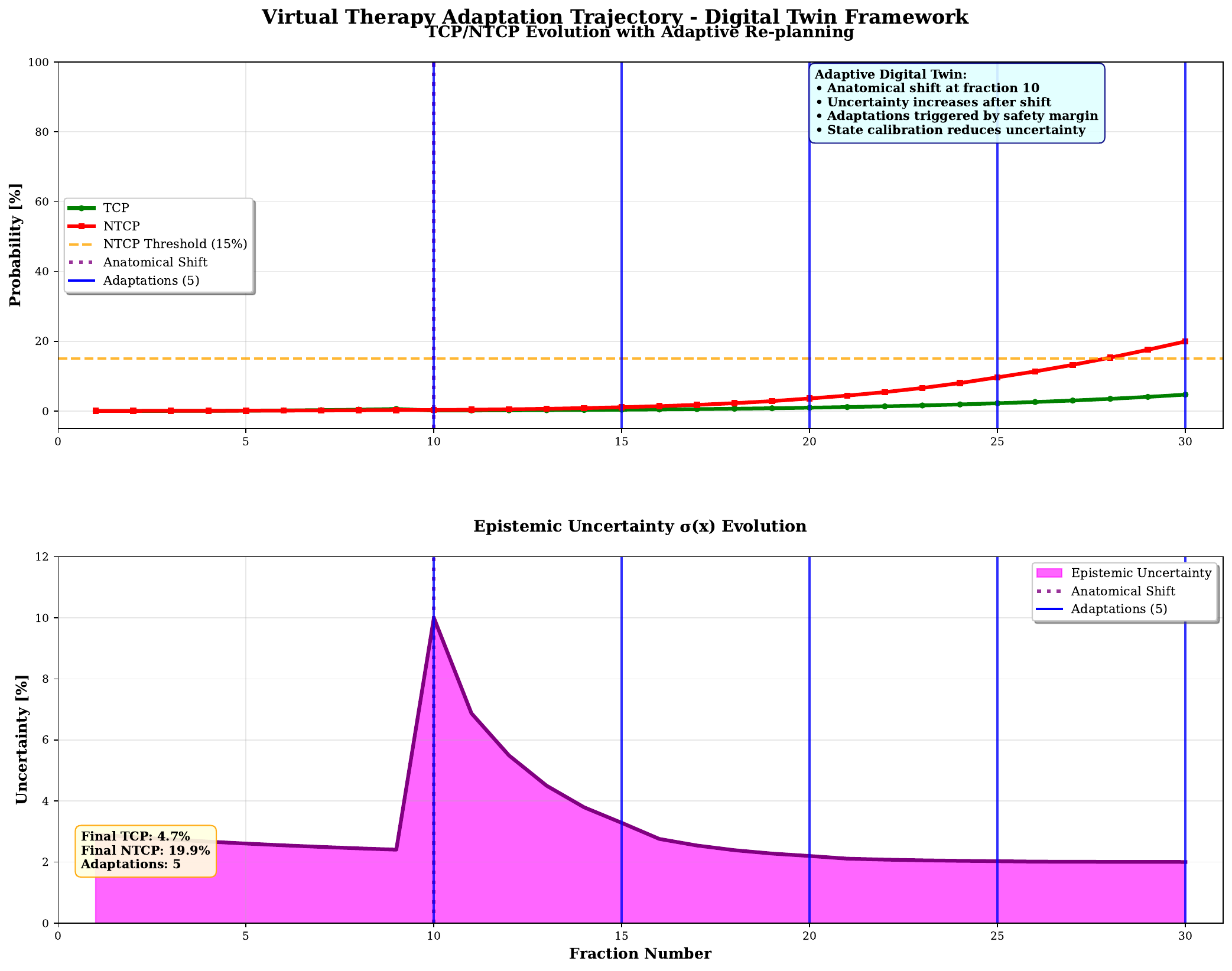}
\caption{Illustrative proxy virtual-therapy adaptation trajectory. Top: surrogate TCP and NTCP objectives across 30 fractions with a simulated anatomical shift at fraction~10. Bottom: corresponding evolution of the epistemic uncertainty summary $\sigma(\mathbf{x})$. The example demonstrates the response of the calibration and decision modules in a proxy experiment; it is not intended as a validated estimate of clinical outcome probabilities.}
\label{fig:adapt_trajectory}
\end{figure*}

Because \openkbp{} is a static planning benchmark and does not provide repeat imaging or delivered-dose logs, fraction-by-fraction adaptation is demonstrated through a proxy virtual-therapy experiment rather than a clinically validated longitudinal cohort analysis. Figure~\ref{fig:adapt_trajectory} illustrates the intended closed-loop behavior of the framework after a simulated anatomical shift at fraction~10. Following the shift, the uncertainty proxy rises and the decision module triggers recalibration and proxy adaptation. In the illustrated run, the updates reduce epistemic uncertainty but do not fully return the surrogate NTCP trajectory below the nominal 15\% reference line by the end of the course. This is worth reporting directly because it shows both the promise and the present limitation of the current action set: the framework can react to rising uncertainty and worsening projected risk, but the present proxy controller should still be interpreted as a proof of concept rather than a clinically validated adaptive strategy~\cite{Li2012TG166}.

\section{Discussion}
The central claim of this paper is that a useful digital twin for radiation-based imaging and therapy must do more than predict dose accurately. It must connect prediction, uncertainty, updating, and decision-making within a single framework. By separating PatientData, Model, Solver, Calibration, and Decision, the proposed architecture makes it possible to improve individual components without losing the logic of the overall closed loop.

The \openkbp{} case study shows both the value and the present boundary of this approach. On one hand, the prototype provides a reproducible open-data implementation of uncertainty-aware dose prediction, proxy recalibration, and virtual-therapy evaluation. On the other hand, \openkbp{} remains a static planning benchmark, so the calibration loop demonstrated here is necessarily an approximation rather than a direct replay of adaptive treatment using delivered-dose logs and repeat imaging. The present manuscript therefore does not claim a fully validated clinical digital twin. Instead, it offers a concrete pathway by which a benchmark dose-surrogate workflow can be extended toward a genuinely actionable digital-twin framework.

Several limitations follow from that scope. The sequential update step is a proxy mechanism based on decoder-only transfer learning rather than a full longitudinal update driven by on-treatment observations. Monte Carlo dropout provides a practical estimate of epistemic uncertainty, but it does not resolve the broader problem of calibration under anatomical change or distribution shift. TCP/NTCP are used as comparative utilities for ranking candidate actions, not as calibrated clinical outcome probabilities for this dataset. Future work should therefore proceed along two directions: stronger technical evaluation on the open benchmark, including broader ablations and stronger uncertainty diagnostics, and translational validation on institutional longitudinal cohorts with delivered-dose logs, repeat imaging, and outcome-linked recalibration.

\section{Conclusion}
This paper presented an actionable digital-twin framework for radiation-based imaging and therapy together with a reproducible \openkbp{}-based prototype. The formulation combines a volumetric dose surrogate, Monte Carlo-dropout uncertainty quantification, proxy sequential recalibration, and constrained decision logic within a single modular workflow. In this way, it shows how an open benchmark for dose prediction can be extended toward the core requirements of a digital twin: updating, uncertainty awareness, and actionability. The present results, including cohort-wide \openkbp{} benchmark evaluation, should still be interpreted at the level of a reproducible prototype rather than a completed clinical system, but they provide a technically grounded foundation for future institutional studies of adaptive and individualized radiation therapy.

\section*{Acknowledgment}
All authors declare that they have no known conflicts of interest in terms of competing financial interests or personal relationships that could have had an influence or are relevant to the work reported in this paper.

\balance
\bibliographystyle{IEEEtran}
\bibliography{references}

\end{document}